\documentclass[aps,prl,twocolumn,superscriptaddress]{revtex4}

\usepackage{graphicx}

\newcommand{\unit}[1] {\ensuremath {~\mathrm {#1} }}
\newcommand{\un}  [1]{\ensuremath{\,\mathrm{#1}}}
\newcommand{\icmax}   {\ensuremath {I_c^{\mathrm{max}}}}

\newcommand{\bc}      {\ensuremath{\beta_C}}
\newcommand{\bl}      {\ensuremath{\beta_L}}
\newcommand{\jphi}    {\ensuremath{\jmath_\phi}}
\newcommand{\jphidot} {\ensuremath{\jmath_{\dot\phi}}}
\newcommand{\half}   {\ensuremath{\frac{1}{2}}}

\newcommand{\pderl}[2]{\ensuremath{\partial #1/\partial #2}}
\newcommand{\real}{\text{Re}}
\newcommand{\imag}{\text{Im}}

\begin{document}


\title{Tunable backaction of a dc SQUID on an integrated micromechanical resonator}

\author{M.~Poot}
\email{m.poot@tudelft.nl}
\affiliation{Kavli Institute of
Nanoscience, Delft University of Technology, Post Office Box
5046, 2600 GA Delft, Netherlands}
\author{S.~Etaki}
\affiliation{Kavli Institute of Nanoscience, Delft University of
Technology, Post Office Box 5046, 2600 GA Delft, Netherlands}
\affiliation{NTT Basic Research Laboratories, NTT Corporation,
Atsugi-shi, Kanagawa 243-0198, Japan}
\author{I.~Mahboob}
\author{K.~Onomitsu}
\author{H.~Yamaguchi}
\affiliation{NTT Basic Research Laboratories, NTT Corporation,
Atsugi-shi, Kanagawa 243-0198, Japan}
\author{Ya.~M.~Blanter}
\author{H.~S.~J.~van~der~Zant}
\affiliation{Kavli Institute of Nanoscience, Delft University of
Technology, Post Office Box 5046, 2600 GA Delft, Netherlands}
\email{h.s.j.vanderzant@tudelft.nl}

\date{\today}

\begin{abstract}
We have measured the backaction of a dc superconducting quantum
interference device (SQUID) position detector on an integrated 1
MHz flexural resonator. The frequency and quality factor of the
micromechanical resonator can be tuned with bias current and
applied magnetic flux. The backaction is caused by the Lorentz
force due to the change in circulating current when the resonator
displaces. The experimental features are reproduced by numerical
calculations using the resistively and capacitively shunted
junction (RCSJ) model.
\end{abstract}


\maketitle

It has recently been demonstrated that a macroscopic mechanical
resonator can be put in a quantum state
\cite{oconnell_nature_quantum_piezo_resonator} by coupling it to
another quantum system. At the same time, linear detectors coupled
to mechanical resonators are rapidly approaching the quantum limit
on position detection. This limit implies that the resonator
position cannot be measured with arbitrary accuracy, as the
detector itself affects the resonator position \cite{caves_RMP}.
This is an example of backaction. Backaction does not just impose
limits, it can also work to one's advantage: Backaction can cool
the resonator, squeeze its motion, and couple and synchronize
multiple resonators. Different backaction mechanisms have been
identified: When using optical interferometers
\cite{gigan_nature_cavity, arcizet_nature_cavity,
thompson_nature_cavity_membrane, anetsberger_natphys_toroid_SiN}
or electronic resonant circuits
\cite{teufel_PRL_stripline_cooling, brown_PRL_circuit_cooling,
rocheleau_nature_stripline_cooling}, backaction results from
radiation pressure. In single-electron transistors (SET)
\cite{naik_nature}, Cooper-pair boxes \cite{lahaye_nature_qubit},
carbon nanotube quantum dots \cite{steele_science_strong_coupling,
lassagne_science_coupled_cnt}, or atomic and quantum point
contacts \cite{flowers_PRL_APC, stettenheim_nature_QPC} backaction
is due to the tunneling of electrons. Recently, we have used a dc
SQUID as a sensitive detector of the position of an integrated
mechanical resonator \cite{etaki_NP_squid_position_detector}. This
embedded resonator-SQUID geometry enables the experimental
realization of a growing number of theoretical proposals for which
a good understanding of the backaction is required
\cite{zhou_PRL_proposal, blencowe_PRB_quantumanalysis,
xue_PRB_two_mode_coupled, huo_APL_squeeze,
nation_PRB_squid_cavity, buks_RPB_rf_decoherence,
xue_NJP_controlable_coupling}.
\begin{figure}[tb]
\includegraphics{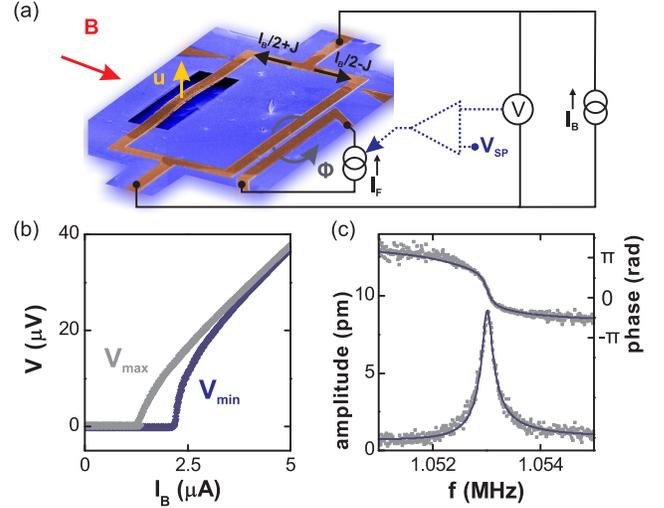}%
\caption{ (a) Schematic overview of the dc SQUID with the
suspended beam and measurement setup. A magnetic field $B$
transduces a beam displacement $u$ into a change in magnetic flux.
A bias current $I_B$ is sent through the SQUID and its output
voltage is measured. The flux $\Phi$ is fine-tuned with a
stripline current $I_F$ that is controlled by a feedback circuit
(dashed) that keeps the output voltage $V$ at $V_{SP}$. (b) The
bias current dependence of the measured $V_\mathrm{min}$ and
$V_\mathrm{max}$. (c) The amplitude (bottom) and phase (top)
response. The line is a fitted harmonic oscillator response
\cite{crosstalk}. \label{fig:overview}}
\end{figure}

In this Letter, we present experiments that show that the dc SQUID
detector exerts backaction on the resonator. By adjusting the bias
conditions of the dc SQUID the frequency and damping of the
mechanical resonator change. The backaction by the dc SQUID has a
different origin than in the experiments mentioned above: It is
due to the Lorentz force generated by the circulating current.
Numerical calculations using the RCSJ model for the dc SQUID
\cite{squidhandbook} reproduce the experimental features.

The device (Fig. \ref{fig:overview}a) consists of a dc SQUID with
proximity-effect-based junctions
\cite{etaki_NP_squid_position_detector}. A part of one arm is
underetched, forming a $1 \unit{MHz}$ flexural resonator with
length $\ell = 50 \unit{\mu m}$. In this Letter we present data on
a device in an in-plane magnetic field of $B = 100 \unit{mT}$.
Measurements have been performed at several magnetic fields and on
an additional device; the observed backaction is similar
\cite{suppinfo}. First the dc SQUID is characterized. The output
voltage of a dc SQUID depends on the magnetic flux through its
loop $\Phi$ \cite{squidhandbook} and we measure the minimum and
maximum voltage ($V_\mathrm{min}$ and $V_\mathrm{\max}$) by
sweeping the flux over a few flux quanta $\Phi_0 = h/2e$ with a
nearby stripline (Fig. 1a). This is repeated for different bias
currents to obtain the current-voltage curves shown in Fig.
\ref{fig:overview}b. The maximum critical current is $\icmax =
2.19 \unit{\mu A}$ and the normal-state resistance of the
junctions is $R = 15.6\unit{\Omega}$. After this characterization,
the dc SQUID is operated at a given setpoint voltage $V_{SP}$
using a feedback loop that adjusts the flux via the stripline
current \cite{etaki_NP_squid_position_detector, squidhandbook}.
The feedback loop is used to reduce low-frequency flux noise and
flux drift and has a bandwidth of $\sim 2 \unit{kHz}$, i.e., it
does not respond to the $1 \unit{MHz}$ resonator signal.

The fundamental mode of the flexural resonator is excited using a
piezo element underneath the sample and the displacement of the
beam is detected as follows: The in-plane magnetic field
transduces a displacement of the beam $u$ into a flux change $\sim
\ell B u$, which in turn changes the voltage over the dc SQUID.
This voltage is amplified using a cryogenic high-electron mobility
transistor followed by a room temperature amplifier and then
recorded using a network analyzer. Figure \ref{fig:overview}c
shows the amplitude and phase of the measured response, from which
the resonance frequency $f_R$ and quality factor $Q$ are obtained.

\begin{figure}[tb]
\includegraphics{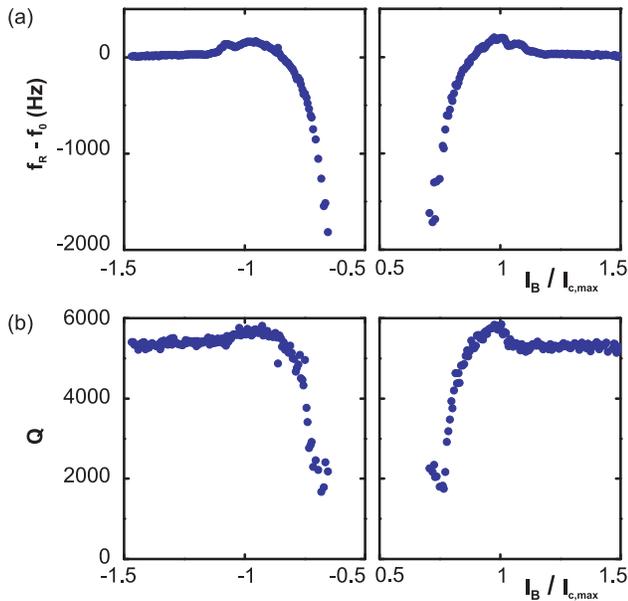}
\caption{Frequency shift (a) and quality factor (b) plotted versus
the normalized bias current. The voltage setpoint was halfway
between $V_\mathrm{min}$ and $V_\mathrm{max}$ in these
measurements. \label{fig:backaction_traces}}
\end{figure}

\begin{figure}[tb]
\includegraphics{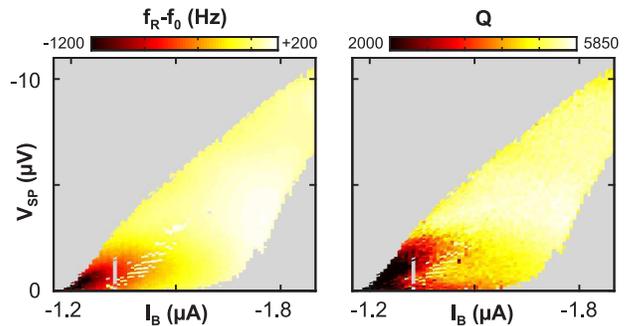}
\caption{Measured bias current and voltage setpoint dependence of
the frequency shift (left) and damping (right). Points without a
good lock are indicated in gray. \label{fig:backaction_maps}}
\end{figure}

To observe backaction of the dc SQUID detector on the resonator,
the frequency response is measured for different bias conditions
of the SQUID. Figure \ref{fig:backaction_traces} shows that both
$f_R$ and $Q$ depend on the bias current $I_B$. The resonance
frequency saturates at $f_0 = 1.053010 \unit{MHz}$ for large
positive and negative bias currents. However, when decreasing the
$I_B$, $f_R$ first goes up by a few hundred Hz around $\icmax$ and
then it decreases rapidly with about -2000 Hz at the lowest stable
setpoint voltage. This is more than $10\times$ the linewidth
$f_R/Q = 194 \unit{Hz}$ of the resonance shown in Fig.
\ref{fig:overview}c. Figure \ref{fig:backaction_traces}b shows
that the quality factor of the resonator changes from $Q_0 = 5300$
to less than 2000. Similar to the resonance frequency, first an
increase and then a stronger decrease in $Q$ is observed when
lowering the bias current.
Figure \ref{fig:backaction_maps} shows that the frequency and
damping can also be changed by adjusting $V_{SP}$, i.e., the flux
through the SQUID loop. The shifts are largest for low setpoints
and low bias currents (dark regions). The regions with a lower
frequency coincide with the regions where the damping has
increased. Bias points with positive frequency shifts and
increases in $Q$ are indicated in white. Finally, by varying the
driving power we confirm that the observed effects are not due to
nonlinearities in the SQUID or in the resonator \cite{suppinfo}.

Unlike for position detectors such as the SET where the backaction
originates from the Coulomb force, the backward coupling between
the SQUID and the beam is the Lorentz force $F_L$
\cite{blencowe_PRB_quantumanalysis, nation_PRB_squid_cavity}. This
force is due to the current  that flows through the beam in the
presence of the magnetic field that couples the resonator and the
SQUID. A displacement changes the flux, and this in turn changes
the circulating current in the loop $J$ \cite{squidhandbook},
giving a different force on the beam. In addition, resonator
motion yields a time-varying flux through the loop, which induces
an electro-motive force and thereby also generates currents that
change the Lorentz force
\cite{cleland_SAA_external_control_dissipation}.

The displacement of the fundamental out-of-plane flexural mode $u$
is given by \cite{cleland_JAP_noise}:
\begin{equation}
m \ddot u + m\omega_0\dot u/Q_0 + m\omega_0^2 u = F_d(t) +
F_L(t).\label{eq:HO}
\end{equation}
The resonator has a mass $m$, (intrinsic) frequency $f_0 =
\omega_0 /2\pi$ and quality factor $Q_0$. $F_d$ is the driving
force and $F_L = aB\ell (I_B/2+J)$ is the Lorentz force. Here, $a
= (u\ell)^{-1}\int_0^\ell u(x)~\mathrm{d}x \approx 0.9$ for the
fundamental mode, so that also $\partial \Phi/\partial u = a B
\ell$ \cite{etaki_NP_squid_position_detector, cleland_JAP_noise}.
For small amplitudes and low resonator frequencies (much smaller
than the characteristic SQUID frequency $\omega_c = \pi R \icmax /
\Phi_0$), the average circulating current can be expanded in the
displacement and velocity $\dot u$ \cite{suppinfo}:
\begin{equation}
J(u, \dot u) = J_0 + \frac{\partial J}{\partial \Phi} a B \ell u +
\frac{\partial J}{\partial \dot \Phi} a B \ell \dot
u.\label{eq:jtransfer}
\end{equation}
Inserting Eq. (\ref{eq:jtransfer}) into Eq. (\ref{eq:HO}) shows
that the $\partial J / \partial \Phi$ term affects the spring
constant $m\omega_R^2$ and thus $f_R$, whereas the $\partial J /
\partial {\dot \Phi}$ term renormalizes the damping. The shifted resonance
frequency and quality factor are:
\begin{eqnarray}
f_R & = & f_0\left(1-\Delta_f \jphi\right)^{1/2}, \textrm{~with~}
\Delta_f = \frac{a^2B^2\ell^2} {m\omega_0^2} \frac{\icmax}
{2\Phi_0}, \label{eq:fr}
\\
Q & = & Q_0 \frac{f_R}{f_0}  \frac{1}{1- \Delta_Q
\jmath_{\dot\phi} }, \textrm{~with~} \Delta_Q =
\frac{a^2B^2\ell^2} {m\omega_0 R} \frac{Q_0} {2\pi} \label{eq:Q}.
\end{eqnarray}
Here, $\jphi = \partial J / \partial \Phi \times 2\Phi_0/\icmax$
and $\jmath_{\dot\phi} = \partial J / \partial \dot \Phi \times 2
\omega_c \Phi_0/\icmax$ are the scaled flux-to-current transfer
functions \cite{hilbert_JLTP_dynamic_input_impedence}. The former
indicates how much the circulating current changes when the flux
through the ring is altered, whereas the latter quantifies the
effect of a time-dependent flux on the circulating current. These
functions were first studied in the analysis of the dynamic input
impedance of tuned SQUID amplifiers
\cite{hilbert_JLTP_dynamic_input_impedence,
falferi_APL_1997_dynamic_input_impedence} and are intrinsic
properties of the SQUID.

Before looking in more detail at the transfer functions, we first
focus on the coupling. The dimensionless parameters $\Delta_f$
\cite{huo_APL_squeeze} and $\Delta_Q$ characterize the backaction
strength. They contain the term $aB\ell$ squared as both the flux
change and the Lorentz force are proportional to the magnetic
field. This implies that the backaction remains the same when the
direction of the magnetic field is reversed and this is what we
observe experimentally. $\Delta_f$ is proportional to $\icmax$,
whereas the damping induced by the SQUID depends on $R$. By a
careful design of the resonator and SQUID, the backaction
strengths can be tuned over a wide range. Eqs. (\ref{eq:fr}) and
(\ref{eq:Q}) show that the largest backaction occurs for large
flux changes $aB\ell$, low spring constants $m\omega^2_0$, and
large circulating currents, i.e., large $I_0$ and low $R$. For the
device studied in this Letter, we estimate $\Delta_f = 4.1 \times
10^{-4}$ and $\Delta_Q = 2.8 \times 10^{-4}$. Finally, note that
the two coupling parameters are related by $\Delta_Q = \Delta_f
\times Q_0 \omega_0 / \omega_c$.

\begin{figure}[tb]
\includegraphics{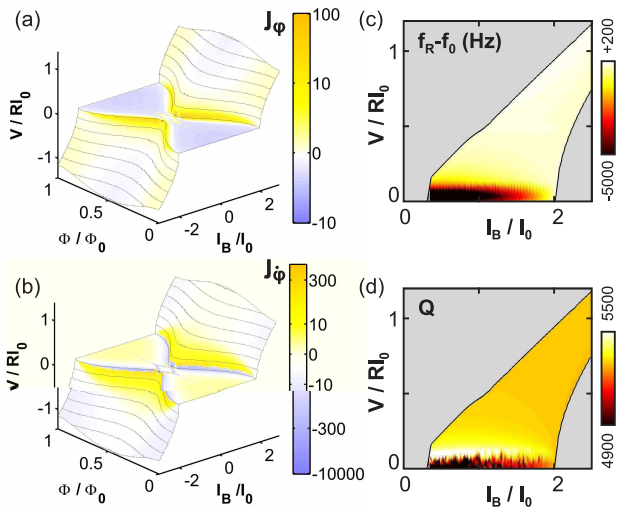}%
\caption{Surface plots with iso-voltage lines at different bias
conditions of a dc SQUID. The (logarithmic) color-scale represents
the calculated flux-to-current transfer functions $\jphi$ (a) and
$\jphidot$ (b). With the values for $\Delta_f$ and $\Delta_Q$, the
frequency shift (c) and quality factor change (d) are calculated.
The simulation is done for the experimental conditions where the
inductive screening parameter $\bl = 0.21$ and the
Stewart-McCumber parameter $\bc = 0.23$ \cite{squidhandbook,
suppinfo}. \label{fig:transfer_function}}
\end{figure}

So far, the analysis did not assume anything about the number of
junctions, nor about their microscopic details. To obtain the
transfer functions $\jphi$ and $\jphidot$, we model the junctions
in the dc SQUID using the RCSJ model \cite{squidhandbook}. The
transfer functions can be calculated analytically in certain
limits \cite{calc_blanter}. However, to obtain their full
bias-condition dependence, $\jphi$ and $\jphidot$ must be
calculated numerically. This is done by simulating the dynamics of
the SQUID in the presence of a time-varying flux \cite{suppinfo}.
Figure \ref{fig:transfer_function}a shows the bias-dependence of
$\jphi$. In the region where $V = 0$, the circulating current
redistributes the bias current between the two junctions such that
no voltage develops. Here the circulating current is of the order
of $\icmax/2$, which gives $\jphi \sim -1$ (blue). In the
dissipative region ($V \ne 0$), the circulating current is
suppressed. Therefore, the circulating current changes rapidly
close to the edge of the dissipative region. The orange color in
Fig. \ref{fig:transfer_function}a indicates that $\jphi$ is large
and positive near the critical current. The largest downward
frequency shift is expected near a half-integer number of flux
quanta, whereas $\jphi$ vanishes for integer flux. With the value
of the coupling parameter $\Delta_f$ and the resonance frequency
$f_0$ the frequency shift is calculated as shown in Fig.
\ref{fig:transfer_function}c.  The maximum value $\jphi = 53$
gives a frequency shift of  $-12 \unit{kHz}$ in the lower-left
corner, which has to be compared with the experimental value of
$\sim -2 \unit{kHz}$. Increasing the bias current above the
critical current results in a smaller $\jphi$ (light yellow and
light blue) that depends linearly on the inductive screening
parameter $\bl$ \cite{squidhandbook} for the experimental
conditions. In this region the simulations predict both positive
(blue) and negative (yellow) value for $\jphi$. Positive and
negative shifts are also observed in the experiment (Figs.
\ref{fig:backaction_traces} and \ref{fig:backaction_maps}). The
largest negative value found in the simulations of that region is
$\jphi = -0.65$, which results in an increase in $f_R$ of $\sim
140 \unit{Hz}$, which is in agreement with the observed value of
$\sim 200 \unit{Hz}$ (Fig. \ref{fig:backaction_traces}a). For even
larger bias currents, the frequency shift vanishes
 $f_R \approx f_0$. This corresponds to the flat regions in Fig.
\ref{fig:backaction_traces}a.

The change in damping is determined by $\jphidot$ as indicated by
Eq. (\ref{eq:Q}). Its dependence on the bias conditions is shown
in Fig. \ref{fig:transfer_function}b. Well inside the
experimentally inaccessible non-dissipative region $\jphidot \sim
+1$ and the backaction results in a small increase in $Q$. In the
opposite limit of large bias currents $\jphidot = -\pi$ (light
blue). This value combined with the small value of $\Delta_Q$
implies that the quality factor in the flat region in Fig.
\ref{fig:backaction_traces}b is close to the intrinsic Q-factor,
$Q_0$. In this region the small additional damping is due to the
current induced by the time-varying flux $\dot \Phi /2R$, which is
dissipated in the junction resistances \cite{calc_blanter}. This
contribution is well-known from magnetomotive readout of
mechanical resonators
\cite{cleland_SAA_external_control_dissipation}. When lowering the
bias current, $\jphidot$ changes sign and rises to about $+500$.
This reduces the damping and might even lead to instability ($Q <
0$) if $\Delta_Q$ is large enough. This decrease of damping
corresponds to the bumps in Fig. \ref{fig:backaction_traces}b. The
largest observed quality factor $Q = 5800$ corresponds to
$\jphidot = +400$, which is in reasonable agreement with the
simulations. Close to the critical current, $\jphidot$ goes to
large \emph{negative} values leading to an enhanced dissipation.
Figure \ref{fig:transfer_function}d shows that the calculated
Q-factor is indeed lowest near the critical current. In summary,
our model shows that although the coupling strength is small, the
dynamics of the dc SQUID greatly enhances the backaction.

Various interesting effects can be observed when the backaction is
strong. If the resonator and SQUID are strongly coupled, the
resonator temperature is set by the effective bath temperature
\cite{naik_nature, flowers_PRL_APC} of the SQUID. The increased
damping cools the resonator, but the shot noise in the bias
current leads to an increase in the force noise on the resonator,
heating it. The question whether the resonator temperature is
above or below the environmental temperature should be addressed
in future research. Furthermore, the dependence of $f_R$ and $Q$
on the bias conditions allows parametric excitation of the
mechanical resonator by either modulating the flux or the bias
current. This enables squeezing of the thermomechanical noise of
the resonator \cite{rugar_PRL_squeezing, huo_APL_squeeze}.
Finally, if the SQUID contains multiple, nearly identical
mechanical resonators, these are coupled to each other by the
backaction. This, in turn, can synchronize their motion and might
lead to frequency entrainment if higher order terms in Eq.
(\ref{eq:jtransfer}) become significant
\cite{shim_science_coupled_resonators}. These examples are only a
few intriguing possibilities of the rich physics connected to the
backaction that we have described in this Letter.

\begin{acknowledgments}
We thank T. Akazaki, H. Okamoto and K. Yamazaki for help with
fabrication and R. Schouten for the measurement electronics. This
work was supported in part by JSPS KAKENHI (20246064 and
18201018), FOM, NWO (VICI grant), and a EU FP7 STREP project
(QNEMS).
\end{acknowledgments}


\begin{thebibliography}{33}
\expandafter\ifx\csname
natexlab\endcsname\relax\def\natexlab#1{#1}\fi
\expandafter\ifx\csname bibnamefont\endcsname\relax
  \def\bibnamefont#1{#1}\fi
\expandafter\ifx\csname bibfnamefont\endcsname\relax
  \def\bibfnamefont#1{#1}\fi
\expandafter\ifx\csname citenamefont\endcsname\relax
  \def\citenamefont#1{#1}\fi
\expandafter\ifx\csname url\endcsname\relax
  \def\url#1{\texttt{#1}}\fi
\expandafter\ifx\csname
urlprefix\endcsname\relax\def\urlprefix{URL }\fi
\providecommand{\bibinfo}[2]{#2}
\providecommand{\eprint}[2][]{\url{#2}}

\bibitem[{\citenamefont{O'Connell et~al.}(2010)\citenamefont{O'Connell,
  Hofheinz, Ansmann, Bialczak, Lenander, Lucero, Neeley, Sank, Wang, Weides
  et~al.}}]{oconnell_nature_quantum_piezo_resonator}
\bibinfo{author}{\bibfnamefont{A.~D.} \bibnamefont{O'Connell}}
  \bibnamefont{et~al.}, \bibinfo{journal}{Nature}
  \textbf{\bibinfo{volume}{464}}, \bibinfo{pages}{697} (\bibinfo{year}{2010}).

\bibitem[{\citenamefont{Caves et~al.}(1980)\citenamefont{Caves, Thorne, Drever,
  Sandberg, and Zimmermann}}]{caves_RMP}
\bibinfo{author}{\bibfnamefont{C.~M.} \bibnamefont{Caves}}
  \bibnamefont{et~al.},  \bibinfo{journal}{Rev. Mod. Phys.} \textbf{\bibinfo{volume}{52}},
  \bibinfo{pages}{341} (\bibinfo{year}{1980}).

\bibitem[{\citenamefont{Gigan et~al.}(2006)\citenamefont{Gigan, Bohm,
  Paternostro, Blaser, Langer, Hertzberg, Schwab, Bauerle, Aspelmeyer, and
  Zeilinger}}]{gigan_nature_cavity}
\bibinfo{author}{\bibfnamefont{S.}~\bibnamefont{Gigan}}  \bibnamefont{et~al.} ,
  \bibinfo{journal}{Nature} \textbf{\bibinfo{volume}{444}}, \bibinfo{pages}{67}
  (\bibinfo{year}{2006}).

\bibitem[{\citenamefont{Arcizet et~al.}(2006)\citenamefont{Arcizet, Cohadon,
  Briant, Pinard, and Heidmann}}]{arcizet_nature_cavity}
\bibinfo{author}{\bibfnamefont{O.}~\bibnamefont{Arcizet}}  \bibnamefont{et~al.},
  \bibinfo{journal}{Nature} \textbf{\bibinfo{volume}{444}}, \bibinfo{pages}{71}
  (\bibinfo{year}{2006}).

\bibitem[{\citenamefont{Thompson et~al.}(2008)\citenamefont{Thompson, Zwickl,
  Jayich, Marquardt, Girvin, and Harris}}]{thompson_nature_cavity_membrane}
\bibinfo{author}{\bibfnamefont{J.~D.} \bibnamefont{Thompson}}  \bibnamefont{et~al.}, \bibinfo{journal}{Nature}
  \textbf{\bibinfo{volume}{452}}, \bibinfo{pages}{72} (\bibinfo{year}{2008}).

\bibitem[{\citenamefont{Anetsberger et~al.}(2009)\citenamefont{Anetsberger,
  Arcizet, Unterreithmeier, Riviere, Schliesser, Weig, Kotthaus, and
  Kippenberg}}]{anetsberger_natphys_toroid_SiN}
\bibinfo{author}{\bibfnamefont{G.}~\bibnamefont{Anetsberger}} \bibnamefont{et~al.}, \bibinfo{journal}{Nat Phys}
  \textbf{\bibinfo{volume}{5}}, \bibinfo{pages}{909} (\bibinfo{year}{2009}).

\bibitem[{\citenamefont{Teufel et~al.}(2008)\citenamefont{Teufel, Harlow,
  Regal, and Lehnert}}]{teufel_PRL_stripline_cooling}
\bibinfo{author}{\bibfnamefont{J.~D.} \bibnamefont{Teufel}}   \bibnamefont{et~al.},
  \bibinfo{journal}{Phys. Rev. Lett.} \textbf{\bibinfo{volume}{101}},
  \bibinfo{eid}{197203} (pages~\bibinfo{numpages}{4}) (\bibinfo{year}{2008}).

\bibitem[{\citenamefont{Brown et~al.}(2007)\citenamefont{Brown, Britton,
  Epstein, Chiaverini, Leibfried, and Wineland}}]{brown_PRL_circuit_cooling}
\bibinfo{author}{\bibfnamefont{K.~R.} \bibnamefont{Brown}}  \bibnamefont{et~al.}, \bibinfo{journal}{Phys. Rev. Lett.}
  \textbf{\bibinfo{volume}{99}}, \bibinfo{eid}{137205}
  (pages~\bibinfo{numpages}{4}) (\bibinfo{year}{2007}).

\bibitem[{\citenamefont{Rocheleau et~al.}(2010)\citenamefont{Rocheleau, Ndukum,
  Macklin, Hertzberg, Clerk, and Schwab}}]{rocheleau_nature_stripline_cooling}
\bibinfo{author}{\bibfnamefont{T.}~\bibnamefont{Rocheleau}}  \bibnamefont{et~al.},
  \bibinfo{journal}{Nature} \textbf{\bibinfo{volume}{463}}, \bibinfo{pages}{72}
  (\bibinfo{year}{2010}).

\bibitem[{\citenamefont{Naik et~al.}(2006)\citenamefont{Naik, Buu, LaHaye,
  Armour, Clerk, Blencowe, and Schwab}}]{naik_nature}
\bibinfo{author}{\bibfnamefont{A.}~\bibnamefont{Naik}}  \bibnamefont{et~al.}, \bibinfo{journal}{Nature}
  \textbf{\bibinfo{volume}{443}}, \bibinfo{pages}{193} (\bibinfo{year}{2006}).

\bibitem[{\citenamefont{LaHaye et~al.}(2009)\citenamefont{LaHaye, Suh,
  Echternach, Schwab, and Roukes}}]{lahaye_nature_qubit}
\bibinfo{author}{\bibfnamefont{M.~D.} \bibnamefont{LaHaye}}  \bibnamefont{et~al.}, \bibinfo{journal}{Nature}
  \textbf{\bibinfo{volume}{459}}, \bibinfo{pages}{960} (\bibinfo{year}{2009}).

\bibitem[{\citenamefont{Steele et~al.}(2009)\citenamefont{Steele, H\"uttel,
  Witkamp, Poot, Meerwaldt, Kouwenhoven, and van~der
  Zant}}]{steele_science_strong_coupling}
\bibinfo{author}{\bibfnamefont{G.~A.} \bibnamefont{Steele}}  \bibnamefont{et~al.}, \bibinfo{journal}{Science}
  \textbf{\bibinfo{volume}{325}}, \bibinfo{pages}{1103} (\bibinfo{year}{2009}).

\bibitem[{\citenamefont{Lassagne et~al.}(2009)\citenamefont{Lassagne,
  Tarakanov, Kinaret, Garcia-Sanchez, and
  Bachtold}}]{lassagne_science_coupled_cnt}
\bibinfo{author}{\bibfnamefont{B.}~\bibnamefont{Lassagne}}  \bibnamefont{et~al.},
  \bibinfo{journal}{Science} \textbf{\bibinfo{volume}{325}},
  \bibinfo{pages}{1107} (\bibinfo{year}{2009}).

\bibitem[{\citenamefont{Flowers-Jacobs
  et~al.}(2007)\citenamefont{Flowers-Jacobs, Schmidt, and
  Lehnert}}]{flowers_PRL_APC}
\bibinfo{author}{\bibfnamefont{N.~E.} \bibnamefont{Flowers-Jacobs}},
  \bibinfo{author}{\bibfnamefont{D.~R.} \bibnamefont{Schmidt}},
  \bibnamefont{and} \bibinfo{author}{\bibfnamefont{K.~W.}
  \bibnamefont{Lehnert}}, \bibinfo{journal}{Phys. Rev. Lett.}
  \textbf{\bibinfo{volume}{98}}, \bibinfo{eid}{096804}
  (pages~\bibinfo{numpages}{4}) (\bibinfo{year}{2007}).

\bibitem[{\citenamefont{Stettenheim et~al.}(2010)\citenamefont{Stettenheim,
  Thalakulam, Pan, Bal, Ji, Xue, Pfeiffer, West, Blencowe, and
  Rimberg}}]{stettenheim_nature_QPC}
\bibinfo{author}{\bibfnamefont{J.}~\bibnamefont{Stettenheim}}  \bibnamefont{et~al.}, \bibinfo{journal}{Nature}
  \textbf{\bibinfo{volume}{466}}, \bibinfo{pages}{86} (\bibinfo{year}{2010}).

\bibitem[{\citenamefont{Etaki et~al.}(2008)\citenamefont{Etaki, Poot, Mahboob,
  Onomitsu, Yamaguchi, and van~der Zant}}]{etaki_NP_squid_position_detector}
\bibinfo{author}{\bibfnamefont{S.}~\bibnamefont{Etaki}}  \bibnamefont{et~al.}, \bibinfo{journal}{Nat Phys}
  \textbf{\bibinfo{volume}{4}}, \bibinfo{pages}{785} (\bibinfo{year}{2008}).

\bibitem[{\citenamefont{Zhou and Mizel}(2006)}]{zhou_PRL_proposal}
\bibinfo{author}{\bibfnamefont{X.}~\bibnamefont{Zhou}} \bibnamefont{and}
  \bibinfo{author}{\bibfnamefont{A.}~\bibnamefont{Mizel}},
  \bibinfo{journal}{Phys. Rev. Lett.} \textbf{\bibinfo{volume}{97}},
  \bibinfo{eid}{267201} (pages~\bibinfo{numpages}{4}) (\bibinfo{year}{2006}).

\bibitem[{\citenamefont{Blencowe and
  Buks}(2007)}]{blencowe_PRB_quantumanalysis}
\bibinfo{author}{\bibfnamefont{M.~P.} \bibnamefont{Blencowe}} \bibnamefont{and}
  \bibinfo{author}{\bibfnamefont{E.}~\bibnamefont{Buks}},
  \bibinfo{journal}{Phys. Rev. B} \textbf{\bibinfo{volume}{76}},
  \bibinfo{eid}{014511} (pages~\bibinfo{numpages}{16}) (\bibinfo{year}{2007}).

\bibitem[{\citenamefont{Xue et~al.}(2007{\natexlab{a}})\citenamefont{Xue,
  xi~Liu, Sun, and Nori}}]{xue_PRB_two_mode_coupled}
\bibinfo{author}{\bibfnamefont{F.}~\bibnamefont{Xue}}  \bibnamefont{et~al.},
  \bibinfo{journal}{Phys. Rev. B} \textbf{\bibinfo{volume}{76}},
  \bibinfo{eid}{064305} (pages~\bibinfo{numpages}{9})
  (\bibinfo{year}{2007}{\natexlab{a}}).

\bibitem[{\citenamefont{Huo and Long}(2008)}]{huo_APL_squeeze}
\bibinfo{author}{\bibfnamefont{W.~Y.} \bibnamefont{Huo}} \bibnamefont{and}
  \bibinfo{author}{\bibfnamefont{G.~L.} \bibnamefont{Long}},
  \bibinfo{journal}{Appl. Phys. Lett.} \textbf{\bibinfo{volume}{92}},
  \bibinfo{eid}{133102} (pages~\bibinfo{numpages}{3}) (\bibinfo{year}{2008}).

\bibitem[{\citenamefont{Nation et~al.}(2008)\citenamefont{Nation, Blencowe, and
  Buks}}]{nation_PRB_squid_cavity}
\bibinfo{author}{\bibfnamefont{P.~D.} \bibnamefont{Nation}},
  \bibinfo{author}{\bibfnamefont{M.~P.} \bibnamefont{Blencowe}},
  \bibnamefont{and} \bibinfo{author}{\bibfnamefont{E.}~\bibnamefont{Buks}},
  \bibinfo{journal}{Phys. Rev. B} \textbf{\bibinfo{volume}{78}},
  \bibinfo{eid}{104516} (pages~\bibinfo{numpages}{17}) (\bibinfo{year}{2008}).

\bibitem[{\citenamefont{Buks and Blencowe}(2006)}]{buks_RPB_rf_decoherence}
\bibinfo{author}{\bibfnamefont{E.}~\bibnamefont{Buks}} \bibnamefont{and}
  \bibinfo{author}{\bibfnamefont{M.~P.} \bibnamefont{Blencowe}},
  \bibinfo{journal}{Phys. Rev. B} \textbf{\bibinfo{volume}{74}},
  \bibinfo{eid}{174504} (pages~\bibinfo{numpages}{8}) (\bibinfo{year}{2006}).

\bibitem[{\citenamefont{Xue et~al.}(2007{\natexlab{b}})\citenamefont{Xue, Wang,
  Sun, Okamoto, Yamaguchi, and Semba}}]{xue_NJP_controlable_coupling}
\bibinfo{author}{\bibfnamefont{F.}~\bibnamefont{Xue}}  \bibnamefont{et~al.},
  \bibinfo{journal}{New J. Phys.} \textbf{\bibinfo{volume}{9}},
  \bibinfo{pages}{35} (\bibinfo{year}{2007}{\natexlab{b}}).

\bibitem[{\citenamefont{Clarke and Braginski}(2004)}]{squidhandbook}
\bibinfo{author}{\bibfnamefont{J.}~\bibnamefont{Clarke}} \bibnamefont{and}
  \bibinfo{author}{\bibfnamefont{A.}~\bibnamefont{Braginski}},
  \emph{\bibinfo{title}{The SQUID Handbook volume 1}}
  (\bibinfo{publisher}{Wiley-VCH Verlag, GmbH and Co. KGaA, Weinheim},
  \bibinfo{year}{2004}).

\bibitem[{sup()}]{suppinfo}
\bibinfo{note}{See EPAPS Document No. [number will be inserted by publisher]
  for supplementary information. For more information on EPAPS, see
  http://www.aip.org/pubservs/epaps.html.}

\bibitem[{\citenamefont{Cleland and
  Roukes}(1999)}]{cleland_SAA_external_control_dissipation}
\bibinfo{author}{\bibfnamefont{A.~N.} \bibnamefont{Cleland}} \bibnamefont{and}
  \bibinfo{author}{\bibfnamefont{M.~L.} \bibnamefont{Roukes}},
  \bibinfo{journal}{Sens. Act. A} \textbf{\bibinfo{volume}{72}},
  \bibinfo{pages}{256} (\bibinfo{year}{1999}).

\bibitem[{\citenamefont{Cleland and Roukes}(2002)}]{cleland_JAP_noise}
\bibinfo{author}{\bibfnamefont{A.~N.} \bibnamefont{Cleland}} \bibnamefont{and}
  \bibinfo{author}{\bibfnamefont{M.~L.} \bibnamefont{Roukes}},
  \bibinfo{journal}{J. Appl. Phys.} \textbf{\bibinfo{volume}{92}},
  \bibinfo{pages}{2758} (\bibinfo{year}{2002}).

\bibitem[{\citenamefont{Hilbert and
  Clarke}(1985)}]{hilbert_JLTP_dynamic_input_impedence}
\bibinfo{author}{\bibfnamefont{C.}~\bibnamefont{Hilbert}} \bibnamefont{and}
  \bibinfo{author}{\bibfnamefont{J.}~\bibnamefont{Clarke}},
  \bibinfo{journal}{J. Low Temp. Phys.} \textbf{\bibinfo{volume}{61}},
  \bibinfo{pages}{237} (\bibinfo{year}{1985}).

\bibitem[{\citenamefont{Falferi et~al.}(1997)\citenamefont{Falferi, Mezzena,
  Vitale, and Cerdonio}}]{falferi_APL_1997_dynamic_input_impedence}
\bibinfo{author}{\bibfnamefont{P.}~\bibnamefont{Falferi}}  \bibnamefont{et~al.},
  \bibinfo{journal}{Applied Physics Letters} \textbf{\bibinfo{volume}{71}},
  \bibinfo{pages}{956} (\bibinfo{year}{1997}).

\bibitem[{cal()}]{calc_blanter}
\bibinfo{note}{Ya. M. Blanter, in preparation.}

\bibitem[{\citenamefont{Rugar and Gr\"utter}(1991)}]{rugar_PRL_squeezing}
\bibinfo{author}{\bibfnamefont{D.}~\bibnamefont{Rugar}} \bibnamefont{and}
  \bibinfo{author}{\bibfnamefont{P.}~\bibnamefont{Gr\"utter}},
  \bibinfo{journal}{Phys. Rev. Lett.} \textbf{\bibinfo{volume}{67}},
  \bibinfo{pages}{699} (\bibinfo{year}{1991}).

\bibitem[{\citenamefont{Shim et~al.}(2007)\citenamefont{Shim, Imboden, and
  Mohanty}}]{shim_science_coupled_resonators}
\bibinfo{author}{\bibfnamefont{S.-B.} \bibnamefont{Shim}},
  \bibinfo{author}{\bibfnamefont{M.}~\bibnamefont{Imboden}}, \bibnamefont{and}
  \bibinfo{author}{\bibfnamefont{P.}~\bibnamefont{Mohanty}},
  \bibinfo{journal}{Science} \textbf{\bibinfo{volume}{316}},
  \bibinfo{pages}{95} (\bibinfo{year}{2007}).

\bibitem[{cro()}]{crosstalk}
\bibinfo{note}{Our fits take the small crosstalk between the driving signal and
  output voltage that is present in the experiment, into account. This effect
  makes the lineshape slightly asymmetric.}

\bibitem[{\citenamefont{Yamaguchi et~al.}(2004)\citenamefont{Yamaguchi,
  Miyashita, and Hirayama}}]{yamaguchi_ASS_qfactor}
\bibinfo{author}{\bibfnamefont{H.}~\bibnamefont{Yamaguchi}},
  \bibinfo{author}{\bibfnamefont{S.}~\bibnamefont{Miyashita}},
  \bibnamefont{and} \bibinfo{author}{\bibfnamefont{Y.}~\bibnamefont{Hirayama}},
  \bibinfo{journal}{Appl. Surf. Sci.} \textbf{\bibinfo{volume}{237}},
  \bibinfo{pages}{645} (\bibinfo{year}{2004}).

\bibitem[{\citenamefont{H\"uttel et~al.}(2009)\citenamefont{H\"uttel, Steele,
  Witkamp, Poot, Kouwenhoven, and van~der Zant}}]{huettel_NL_highQ}
\bibinfo{author}{\bibfnamefont{A.~K.} \bibnamefont{H\"uttel}} \bibnamefont{et~al.},
 \bibinfo{journal}{Nano Lett.}
  \textbf{\bibinfo{volume}{9}}, \bibinfo{pages}{2547} (\bibinfo{year}{2009}).

\bibitem[{\citenamefont{LaHaye et~al.}(2004)\citenamefont{LaHaye, Buu,
  Camarota, and Schwab}}]{lahaye_science}
\bibinfo{author}{\bibfnamefont{M.~D.} \bibnamefont{LaHaye}}\bibnamefont{et~al.},
  \bibinfo{journal}{Science} \textbf{\bibinfo{volume}{304}},
  \bibinfo{pages}{74} (\bibinfo{year}{2004}).

\bibitem[{\citenamefont{Zolfagharkhani
  et~al.}(2005)\citenamefont{Zolfagharkhani, Gaidarzhy, Shim, Badzey, and
  Mohanty}}]{zolfagharkhani_PRB_qfactor}
\bibinfo{author}{\bibfnamefont{G.}~\bibnamefont{Zolfagharkhani}} \bibnamefont{et~al.},
  \bibinfo{journal}{Phys. Rev. B} \textbf{\bibinfo{volume}{72}},
  \bibinfo{eid}{224101} (pages~\bibinfo{numpages}{5}) (\bibinfo{year}{2005}).

\end{thebibliography}

\section*{Temperature and power dependence}
\begin{figure}[tb]
\centering
\includegraphics[width=8.6cm]{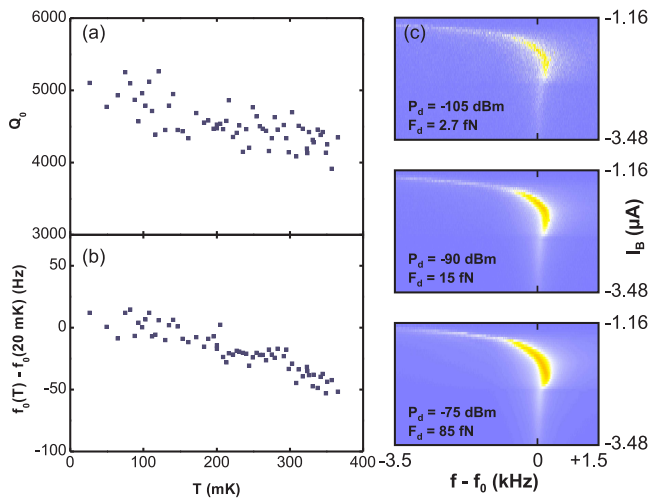}
\caption{Temperature dependence of the intrinsic quality factor
$Q_0$ (a) and resonator frequency $f_0$ (b). (c) Colorscale plot
of the oscillator response at $B = 100 \unit{mT}$ at different
driving powers $P_d$. This measurement is done with a voltage
setpoint halfway between $V_\mathrm{min}$ and $V_\mathrm{max}$.
\label{fig:power}}
\end{figure}

Figures \ref{fig:power}a and b show the temperature dependence of
the resonance frequency and quality factor. These values are
measured at large bias currents where the backaction is
negligible. The frequency change due to temperature is small
compared to the observed backaction (see main text). The intrinsic
quality factor decreases significantly with increasing
temperature. This rules out that the observed frequency shift and
Q-factor change are caused by heating of the resonator due to
Joule heating in the junctions: We observe an \emph{increase} in
quality factor with increasing bias current and voltage setpoint
(i.e. dissipation), but a \emph{decrease} in quality factor with
increasing temperature. An increased damping at higher
temperatures is seen more often in micro- or nanomechanical
resonators \cite{yamaguchi_ASS_qfactor, huettel_NL_highQ,
lahaye_science, zolfagharkhani_PRB_qfactor}.

The observed frequency shift and change in damping do not depend
on the driving power. As shown in Figure \ref{fig:power}c, the
measured resonator response stays the same in all panels. Although
the driving power is changed by three orders of magnitude, the
only effect is that the signal-to-noise ratio becomes better when
increasing the power.

For the highest driving power ($P_d = -75 \unit{dBm}$) and highest
Q-factor ($Q \sim 5800$) the amplitude of the resonator motion is
$u_{\mathrm{max}} = 20 \unit{pm}$, as determined using the
calibrated displacement responsivity
\cite{etaki_NP_squid_position_detector}. The flux through the dc
SQUID is then modulated with an amplitude $a B \ell
u_{\mathrm{max}} = 0.02 \unit{\Phi_0}$. So even for the largest
resonator motion, the change in flux is much smaller than a flux
quantum. Exactly on resonance, the piezo motion is $Q$ times
smaller than $u_{\mathrm{max}}$, about $3 \unit{fm}$. The driving
force $F_d$ is then given by the resonator mass $m$ times the
acceleration of the piezo element, $F_d = m \omega^2 u_p \approx m
\omega_0^2 u_p$. The measurements shown in the main text are done
with a driving power of $- 80 \unit{dBm}$ ($F_d = 48 \un{fN}$).

\section*{Device B}
All effects that we have observed in the device that we discuss in
the main text have also been measured in a second device. The
resonator in this dc SQUID operates around $2 \unit{MHz}$. The
maximum critical current of device B is $\icmax = 2I_0 = 2.4
\unit{\mu A}$ at $B = 115 \unit{mT}$ and $\icmax = 1.0 \unit{\mu
A}$ at $B = 130 \unit{mT}$.

\begin{figure}[tb]
\centering
\includegraphics[width=8.6cm]{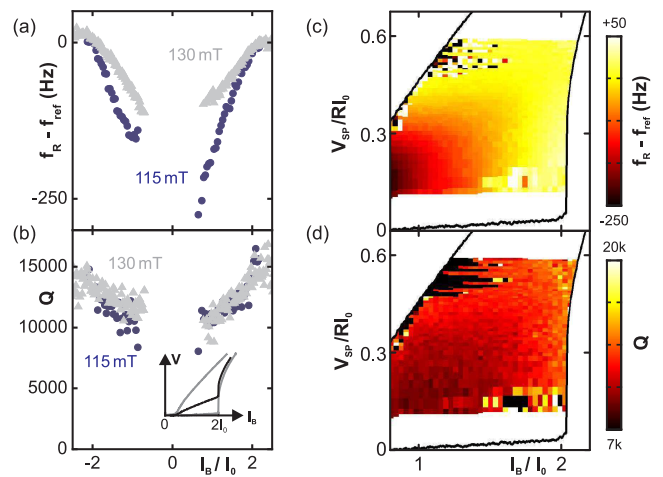}
\caption{Backaction measurements of device B. (a) Frequency shift
and (b) quality factor plotted versus the normalized bias current
at $B = 115 \un{mT}$ (circles) and $B = 130 \un{m}T$ (triangles).
The inset depicts the voltage setpoint for these measurements with
respect to $V_\mathrm{min}$ and $V_\mathrm{max}$. Bias current and
setpoint dependence of the frequency shift (c) and damping (d) at
$B = 115$ mT. The solid lines indicate the measured
$V_\mathrm{min}$ and $V_\mathrm{max}$. \label{fig:devB}}
\end{figure}

In the measurements on device B, the feedback loop could not
maintain the SQUID voltage for low voltage setpoints. Also, a less
sensitive, room temperature amplifier was used for the resonator
signal. Its lower gain resulted in an increased scatter in the
data and the inability to explore the region with the highest
backaction, i.e., at low bias current and low voltage setpoints.
The measurements in Fig. \ref{fig:devB} show qualitatively the
same backaction as the data presented in the main text.

\section*{The RCSJ model for the dc SQUID}
To calculate the backaction, the dc SQUID is modelled using the
resistively- and capacitively-shunted junction (RCSJ) model. This
widely-used model is discussed in detail in Ref.
\cite{squidhandbook}. The introduction to this model presented
here is largely based on this review. A current $I_B$ is sent
through the SQUID and the circulating current $J$ redistributes
the current over the two junctions, which we assume to be
identical. In the RCSJ model, the two junctions (labelled with
$i=1,2$) are modelled as a resistor ($R$), capacitor ($C$) and an
``ideal'' Josephson junction with critical current $I_{0}$ in
parallel. The voltage over each junction is related to the time
derivative of the phase difference $\delta_i$ of the
superconducting wave function: $V_i = \Phi_0 \ \dot \delta_i/
2\pi$, where $\Phi_0 = h/2e = 2.05 \times 10^{-15} \unit{T m^2}$
is the flux quantum. Current conservation yields two second-order
differential equations, governing the time-dependence of the phase
differences $\delta_{1,2}$ of the junctions:
\begin{eqnarray}
\frac{\Phi_0}{2\pi} C \ddot{\delta}_{1} + \frac{\Phi_0}{2\pi}
\frac{1}{R}\dot{\delta}_{1} + I_{0} \sin \delta_{1} & = & \half
I_{B} + J,\\
\frac{\Phi_0}{2\pi} C \ddot{\delta}_{2} + \frac{\Phi_0}{2\pi}
\frac{1}{R}\dot{\delta}_{2} + I_{0} \sin \delta_{2} & = & \half
I_{B} - J.\end{eqnarray}
These equations are coupled to each other by the amount of flux
piercing the loop:
\begin{equation}
\delta_2 - \delta_1 = 2\pi\cdot\Phi_{\mathrm{tot}}/\Phi_0.
\end{equation}
The total flux $\Phi_{\mathrm{tot}}$ has two contributions: the
externally applied flux $\Phi$ (which also includes the flux due
to the resonator displacement) and the flux due to the circulating
current flowing through the inductance of the loop $L$, i.e.,
$\Phi_{\mathrm{tot}} = \Phi + LJ$.

The equations are scaled to simplify their numerical integration.
This yields:
\begin{eqnarray}
\bc \ddot{\delta_1} + \dot{\delta_1} + \sin \delta_1 &=& \imath_B/2+\jmath \\
\bc \ddot{\delta_2} + \dot{\delta_2} + \sin \delta_2 &=& \imath_B/2-\jmath \\
2\pi (\phi + \bl \jmath/2) & = & \delta_2 - \delta_1 .
\end{eqnarray}
The bias current and circulating current are normalized using the
critical current: $\imath_B = I_B/I_0$ and $\jmath = J/I_0$.
Furthermore, time is scaled using the characteristic frequency
$\omega_c = 2\pi RI_0/\Phi_0$, fluxes using the flux quantum,
i.e., $\phi = \Phi/\Phi_0$, and voltages using the characteristic
voltage $RI_0$ so that $v = V/RI_0 = (\dot \delta_1+\dot
\delta_2)/2$. The parameter $\bl = 2I_0L/\Phi_0$ and $\bc = 2\pi
I_0 R^2 C/\Phi_0 $ are the inductive screening parameter and the
Stewart-McCumber number respectively. The inductive screening
parameter indicates how much a change in flux is screened by the
circulating current $J$ flowing through the self-inductance of the
loop $L$, whereas $\bc$ indicates the importance of inertial terms
due to the junction capacitance $C$.

\begin{figure}[tb]
\centering
\includegraphics{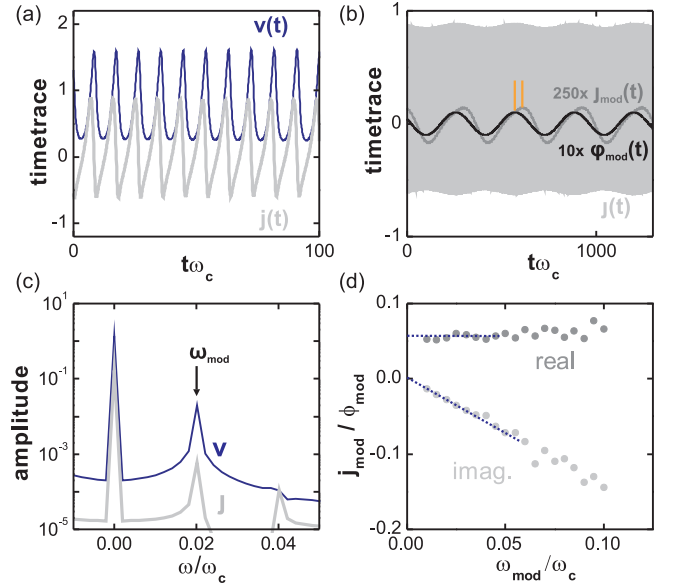}
\caption{(a) Calculated timetrace of the circulating current
$\jmath(t)$ and voltage $v(t)$ in the absence of noise. (b)
Circulating current for a small ($\phi_\mathrm{mod}= 0.01$)
modulation of the flux with $\omega_\mathrm{mod} = 0.02$. The
time-averaged value of the circulating current $\overline \jmath
(t)$ has a phase shift with respect to the modulation $\phi(t)$ as
indicated by the orange lines. (c) Absolute value of the Fourier
transform of the timetraces of the SQUID voltage and circulating
current shown in (b). (d) The modulation frequency dependence of
the real (dark gray) and imaginary (light gray) parts of the
transfer function. The lines are a guide to the eye. These
simulations were done for the dc SQUID parameters from the main
text ($\bl = 0.21$ and $\bc = 0.23$) at $\imath_B = 2$ and $\phi =
0.25$. \label{fig:simulation}}
\end{figure}

The three equations are integrated numerically for different bias
conditions, i.e., different values for the bias current $I_B$ and
for the flux $\Phi$ through the SQUID. Figure
\ref{fig:simulation}a shows typical examples of calculated
time-traces of the circulating current $\jmath$ and voltage $v$.
Both are rapidly oscillating at a frequency of $0.69~\omega_c$,
which is the Josephson frequency that equals the average value of
the voltage $\overline v$ \cite{squidhandbook}. The maximum
current that the dc SQUID can carry without generating a voltage
equals the sum of the critical currents of the two junctions:
$\icmax = 2I_{0}$.

The critical current ($I_0 = 1.1 \unit{\mu A}$) and the
normal-state resistance ($R = 15.6 \unit{\Omega}$) of the
junctions are estimated from the IV-characteristics (Fig. 1b of
the main text). Using finite-element simulations we estimate $L =
175 \unit{pH}$ for our device. Finally, the capacitance $C = 0.6
\unit{pF}$ is obtained from the position of the LC resonance in
the dc SQUID \cite{squidhandbook}.

\section*{Calculation of the transfer functions}
When the resonator moves, the flux through the dc SQUID loop is
altered, which in turn changes the average circulating current
$J$. In principle, $J(t)$ could depend on the all the past
displacements, $u(t')$ for $t' < t$. However, the dc SQUID reacts
at a frequency ($\sim \omega_c/2\pi \sim 8 \un{GHz}$) that is much
faster than that of the resonator ($1 \un{MHz}$), so the
circulating current $J(t)$ is expected to depends on the
\emph{instantaneous} displacement $u(t)$. For small amplitudes ($u
\ll \Phi_0/aB\ell$) the response is linear and gives a
contribution $J_1(t) = c_1 u(t)$. Another contribution comes from
the velocity of the resonator, $\dot u $, which causes a
time-varying flux. This generates an electromotive force in the
SQUID loop (Faraday's induction law), which also changes the
circulating current by $J_2(t) = c_2 \dot u (t)$. Combining these
two effects gives $J(t) = J(u,\dot u ) = c_1 u(t) + c_2 \dot
u(t)$. The values of the parameters $c_1$ and $c_2$ depend on the
dynamics of the dc SQUID and are $c_1 = \pderl{J}{u}$ and $c_2 =
\pderl{J}{\dot u}$. As discussed in the main text, these
quantities are related to the intrinsic flux-to-current transfer
functions, $\jphi = \Phi_0 / (a B \ell I_0) \times \pderl{J}{u}$
and $\jphidot = \omega_c \Phi_0 / (a B \ell I_0) \times
\pderl{J}{\dot u}$.

In principle $\jmath_\phi$ can be obtained by calculating the
average circulating current at different fluxes and then
numerically differentiating this to obtain $\jmath_\phi =
\partial \jmath / \partial \phi$. However, for the
velocity-dependent transfer function this is not possible. Our
method for calculating these transfer functions $\jphi$ and
$\jphidot$ works as follows: We calculate the steady-state
response of the circulating current with a small modulation added
to the applied flux, $\phi \rightarrow \phi + \phi_\mathrm{mod}
\cos(\omega_\mathrm{mod}t)$. Figure \ref{fig:simulation}b shows
that this modulates the circulating current $\jmath(t)$. Figure
\ref{fig:simulation}c shows the Fourier transform of the
circulating current and the voltage. In the spectrum of both
$\jmath$ and $v$ a peak appears at the modulation frequency. The
real part of the peak corresponds to the derivative $\jmath_\phi =
\real[\jmath_\mathrm{mod}/\phi_\mathrm{mod}]$, while
$\imag[\jmath_\mathrm{mod}/\phi_\mathrm{mod}] = -
\omega_\mathrm{mod} \jmath_{\dot \phi}$ and similar for
$v_\mathrm{mod}$. The frequency dependence in Fig
\ref{fig:simulation}d, shows a constant
$\real[\jmath_\mathrm{mod}]$ and a linearly increasing
$\imag[\jmath_\mathrm{mod}]$ as indicated by the dotted lines. The
transfer functions do not depend on the modulation frequency,
provided that it is sufficiently low. This confirms that the
circulating current only depends on the instantaneous displacement
and velocity as was postulated at the beginning of this Section.

\end{document}